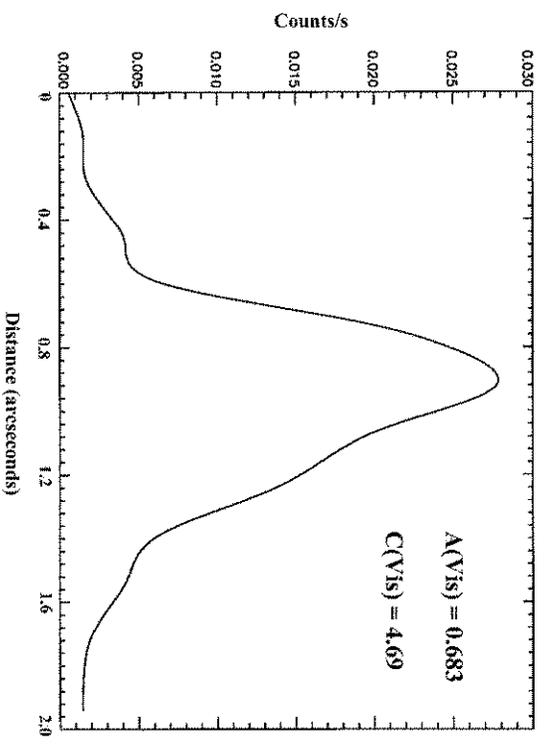

4-916.0

A(Vis) = 0.124
C(Vis) = 3.75

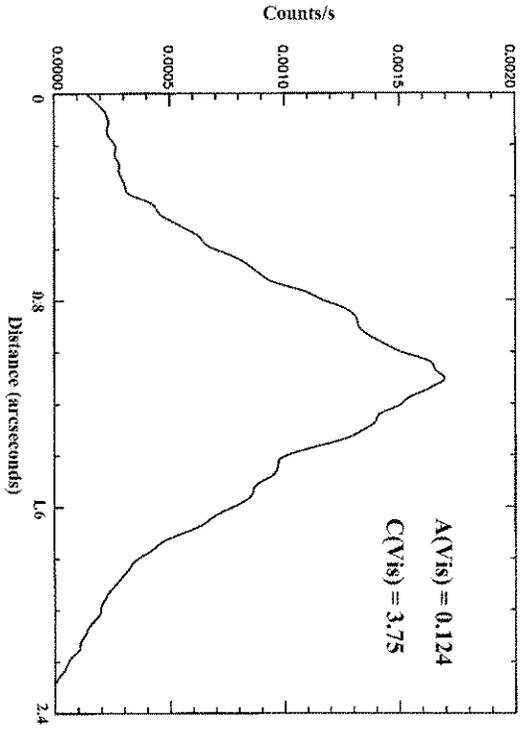

3-430.1

A(Vis) = 0.683
C(Vis) = 4.69

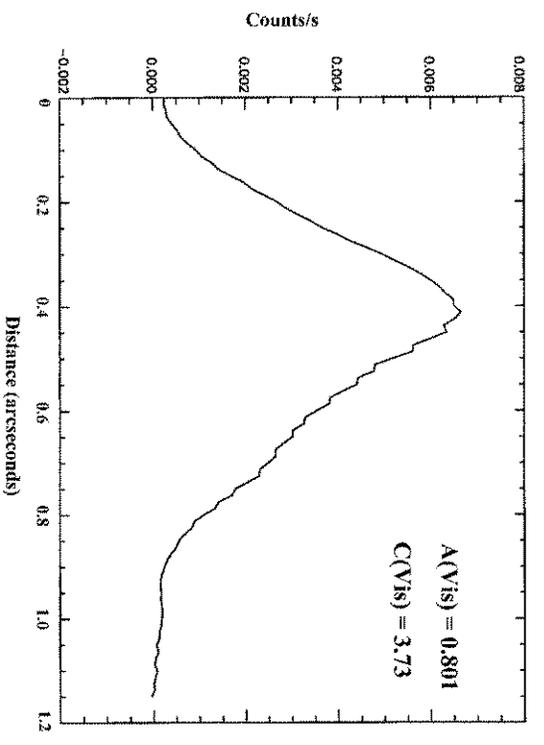

4-660.0

A(Vis) = 0.801
C(Vis) = 3.73

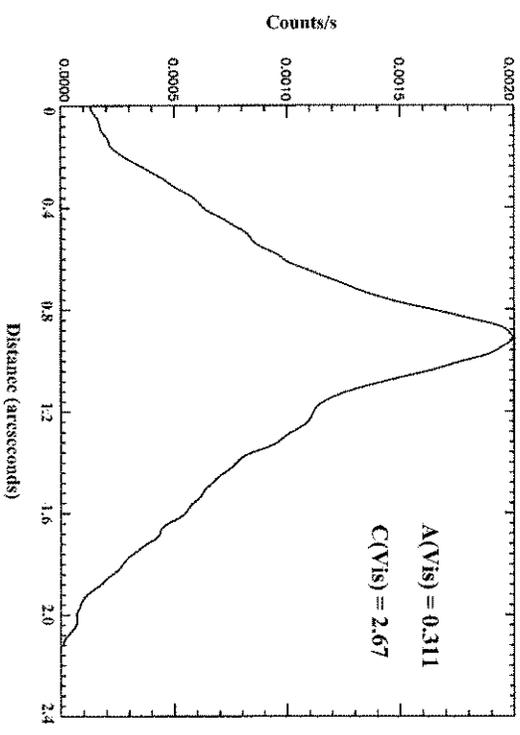

4-558.0

A(Vis) = 0.311
C(Vis) = 2.67

(b)

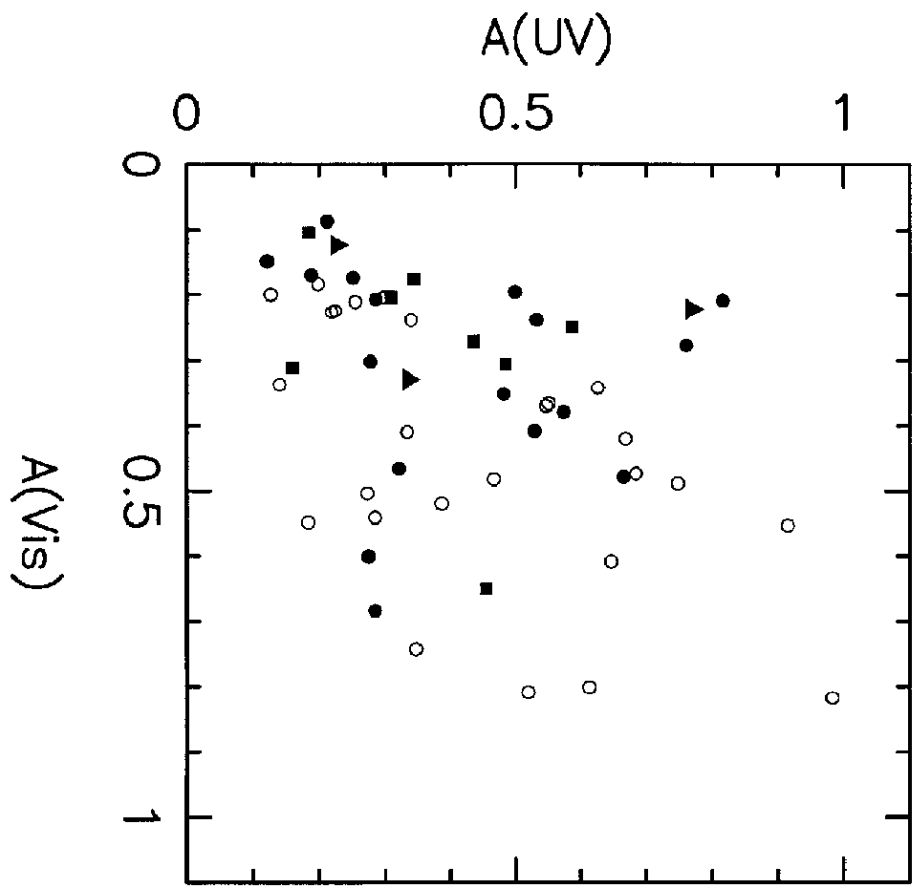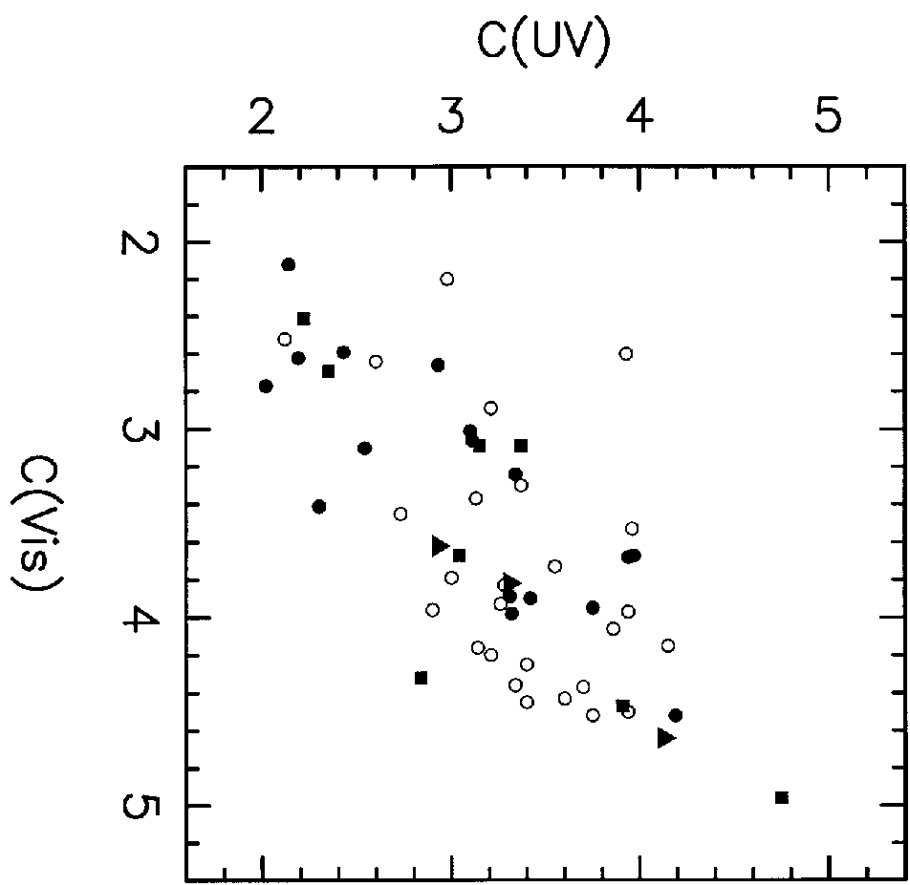

# A Multivariate Analysis of Galaxies

# in the Hubble Deep Field North

Michael R. Corbin

Andrea Urban

Elizabeth Stobie

Rodger I. Thompson

&

Glenn Schneider

NICMOS Group, Steward Observatory, The University of Arizona,

Tucson, AZ  85721; contact: mcorbin@as.arizona.edu

ABSTRACT

We use the ultraviolet and optical WFPC2 and near-infrared NICMOS images of the Hubble Deep Field North to measure and statistically compare an array of parameters for over 250 of the galaxies it contains. These parameters include redshift, rest-frame visible asymmetry and concentration, bolometric luminosity and extinction-corrected star formation rate. We find only one strong correlation, between bolometric luminosity and star formation rate, from which early-type galaxies noticeably deviate. When our asymmetry measurements are combined with those of a sample of nearby galaxies covering the full Hubble sequence, we find a weak correlation between redshift and rest-frame visible asymmetry, consistent with the qualitative evidence of galaxy morphological evolution from these and other deep *Hubble Space Telescope* images. The mean values of these asymmetry measurements show a monotonic increase with redshift interval over the range $0 < z < 2$, increasing by a factor of approximately three. If this trend is real, it suggests that galaxy morphological evolution within the last ~ 70% of the Hubble time is a *gradual* process that is continuing through the present cosmological epoch. There is evidence that the dominant source of this evolution is the "minor" mergers of disk galaxies with smaller companions, which could also transform late-type spiral galaxies to early-type spiral galaxies. Interestingly, in contrast to local galaxies we find no correlations between galaxy star formation rate and either UV or visible asymmetry. This could arise if the star formation of high-redshift galaxies proceeds in episodes that are short (~ 100 Myr) relative to the time scales over which galaxy mergers produce strong asymmetries (~ 500 Myr), a result suggested by the high star formation rates of Lyman break galaxies.

Subject headings: cosmology: observations -- galaxies: evolution -- galaxies: structure -- galaxies: fundamental parameters

1. INTRODUCTION

A major achievement of the *Hubble Space Telescope* (HST) has been the discovery that galaxies at intermediate and high redshift are often interacting, paired, or are otherwise morphologically peculiar (see e.g. Burkey et al. 1994; Glazebrook et al. 1995; Abraham et al. 1996a, 1996b; Driver et al. 1998; Im et al. 1999; van den Bergh et al. 2000). Concern that this apparent peculiarity is the result of viewing galaxies above $z \sim 1$ in the rest-frame ultraviolet, in which their emission is dominated by clusters of OB stars, has largely been alleviated by images from the Near Infrared Camera and Multi-Object Spectrometer (NICMOS) instrument, which provide a "morphological *K*-correction" for galaxies above such redshifts by covering their rest-frame visible emission. NICMOS images of the Hubble Deep Field North (HDFN), in combination with the visible images obtained by the Wide Field Planetary Camera 2 (WFPC2), have revealed a generally good (but not uniform) correspondence between the morphologies of $z > 1$ galaxies in rest-frame ultraviolet and visible portions of their continua (Bunker 1999; Dickinson 1999). Corbin et al. (2000a) also find a larger fraction of peculiar/interacting galaxies at intermediate redshift in the NICMOS parallel fields than is observed in the local universe. In addition to this increase in morphological peculiarity with redshift is the evidence of a concurrent increase in the fraction of photometric and kinematic galaxy pairs (e.g. Neuschaefer et al. 1997; Le Fevre et al. 2000). These results present strong evidence that galaxies have undergone significant structural evolution within the last ~ 2/3 of the Hubble time, an evolution in which merging appears to play a key role, and generally support hierarchical models of galaxy formation (e.g. Baugh et al. 1998; Kaufmann et al. 1999).

Galaxy star formation rates also show a strong increase with redshift, at least to $z \cong 1.5$ (e.g. Madau, Pozzetti & Dickinson 1998; Thompson, Weymann & Storrie-Lombardi 2000, hereafter TWSL). While the form of this evolution above $z \cong 1.5$ remains highly uncertain due to the role of dust obscuration and selection effects (see TWSL and references therein), it apparently proceeds concurrently with the aforementioned morphological evolution. Similarly, this morphological evolution should at some level be accompanied by an evolution in galaxy luminosity and size. However, the aforementioned studies have generally concentrated on the change in only one of these quantities with redshift, rather than investigating their interrelationships. Galaxy morphological evolution has also not been properly quantified: The existing evidence of it is based on qualitative and subjective classifications of the galaxies in the HDFN and other deep HST images (e.g. Im et al. 1999; van den Bergh et al. 2000). Abraham et al. (1996a) attempt such a quantification using measurements of the asymmetries and concentrations of HDFN galaxies in the *I* band image, but their results are limited by the lack of redshifts available for the objects at the time of their study, and by the associated lack of a morphological *K*-correction provided by NICMOS images for the $z > 1$ galaxies in the field.

These issues have motivated us to perform a statistical analysis of the galaxies in the HDFN involving all the potentially correlated parameters, and in particular to attempt to quantify their morphological evolution. The NICMOS images of the HDFN importantly extend the coverage of the rest-frame visible emission of its galaxies to $z \sim 3$, thereby allowing their comparison with galaxies at lower redshifts and in the local universe. We specifically seek to investigate the relationships between all relevant quantities, including redshift, asymmetry, concentration, size, bolometric luminosity and star formation rate, using measurements from the combined WFPC2 and NICMOS images of the field. Such an analysis will in principle place more comprehensive constraints on models of galaxy formation and evolution than are currently available, modulo the limitations inherent in the small area of the HDFN. We additionally include in some of our comparisons measurements made for nearby galaxies, in order to extend our analysis to the current cosmological epoch. Our approach is similar to that successfully applied in studies of quasar spectra (e.g. Boroson & Green 1992; Corbin & Boroson 1996), i.e., we test each measured quantity for correlation with every other quantity, and apply a principal components analysis to the resulting correlation matrix as a means of establishing the dominant sources of variance in the sample. In the following section we describe our data and measurements. In § 3 we present the results of our statistical analysis, and conclude with a discussion of them in § 4. All redshift-dependent quantities have been scaled to a cosmology of $H_0 = 65$ km s$^{-1}$ Mpc$^{-1}$, $\Omega_M = 0.3$, $\Omega_\Lambda = 0.7$.

## 2. DATA AND MEASUREMENTS

Our analysis is based on the final HDFN WFPC2 images of Williams et al. (1996) and the NICMOS images of Thompson et al. (1999; hereafter T99) and Dickinson (1999; hereafter D99), the latter retrieved from the HST archive and calibrated in the same way as the T99 images. All images have been dithered and drizzled as described in Williams et al. (1996) and T99. The D99 NICMOS images cover the same area as the WFPC2 images, while the T99 images consist of a single NICMOS Camera 3 field, covering approximately one-fifth of the total WFPC2 area. The T99 images go slightly deeper than the D99 images, and so were used in the area in which these images overlap.

### 2.1 *Luminosities, Star Formation Rates, and Related Quantities*

TWSL have used the T99 images in combination with the portion of the WFPC2 images that they overlap to measure galaxy fluxes (when possible) in each of the six associated filters (F300W, F450W, F606W and F814W of WFPC2, and F110W and F160W of NICMOS, the latter centered at 1.1 μm and 1.6 μm). The addition of the NICMOS fluxes greatly improves constraints on the galaxy spectral energy distributions (SEDs). TWSL have used these new SEDs to measure the object photometric redshifts, bolometric luminosities, and star formation rates using a detailed galaxy spectral template fitting procedure. This procedure uses an array of galaxy SED templates which cover a range from very actively star forming objects to objects dominated by evolved stellar populations. Specifically, starting from a basic set of six templates derived from Coleman, Wu & Weedman (1980), Calzetti, Kinney & Storchi-Bergman (1994) and the 1996 version of the Bruzual & Charlot (1993) SED library, TWSL apply the galaxy dust extinction law of Calzetti, Kinney & Storchi-Bergman (1994) for various amounts of extinction to create an effective array of 51 templates covering a wide range of star formation rates and internal extinctions. A correction for intergalactic extinction to the redshifted templates using the formulation of Madau et al. (1996) is also included. These SEDs are then fitted to the galaxy flux points using a $\chi^2$ procedure to estimate the object redshift. This redshift value is then used to measure the bolometric luminosity by integrating the rest-frame SED of the best-fitting template, after adding back the amount of flux estimated to be lost due to dust extinction. TWSL also estimate the fraction of the bolometric luminosity that is re-radiated at 850 μm from the extinction value of the best-fitting template. The star formation rate is estimated from the unextincted flux at the rest-frame wavelength of 1500 Å using the relation of Madau, Pozzetti & Dickinson (1998). Support for the TWSL SED fitting method is provided by the close agreement between their photometric redshift values of the HDFN galaxies and those obtained spectroscopically (see their Figure 3).

We include all these measurements (photometric redshift, bolometric luminosity, star formation rate and fractional re-emission, denoted as "frac") in our analysis. We also measure the ratio of star formation rate to bolometric luminosity as an additional parameter, given the evidence that the ratio of star formation rate to galaxy mass (often called specific star formation rate) shows a strong inverse correlation with galaxy mass (Guzmán et al. 1997; Brinchmann & Ellis 2000). This correlation suggests that the ratio of star formation rate to bolometric luminosity is also of diagnostic importance. The TWSL measurements are for 276 objects in the T99 NICMOS image of the HDFN; the D99 images are not included in this analysis. For the galaxies covered by the D99 images, we use the photometric redshifts in the current on-line version of the catalog of Fernández-Soto, Lanzetta & Yahil (1999), which are based on combined WFPC2 fluxes and near-infrared fluxes obtained from ground-based observations. For all objects for which spectroscopic redshifts are available in the Fernández-Soto et al. (1999) catalog, these redshifts have been used in place of the photometric redshifts.

## 2.2 *Morphological Parameters and Sizes*

We attempt to quantify the galaxy morphologies using the asymmetry and concentration parameters of Conselice, Bershady & Jangren (2000; hereafter C2000) and Bershady, Jangren & Conselice (2000). The validity of measuring these parameters for high-redshift galaxies in WFPC2 and NICMOS images is discussed by C2000. Briefly, based on CCD images of the local galaxy sample of Frei et al. (1996), C2000 find that asymmetry measurements are not strongly affected by decreases in angular resolution until a resolution limit ~ 1 $h_{75}^{-1}$ kpc, at which point they sharply decrease relative to their initial value (see their Figure 19). The resolution of the drizzled WFPC2 images (approximately 0.04" pix$^{-1}$) is higher than this for the adopted cosmology up to $z \cong 2$, while that of the drizzled NICMOS images (approximately 0.1" pix$^{-1}$) matches this limit more critically for the same redshift range. There thus should not exist any strong systematic errors in the asymmetry measurements of galaxies in either the WFPC2 or NICMOS images, although we return to further discussion of this point in later sections.

The concentration parameter was first defined by Kent (1985), and a similar asymmetry parameter has been used by Abraham et al. (1996a, 1996b). These parameters are defined as

$$A \equiv \frac{\Sigma|(I_0 - I_{180})|}{2\Sigma|(I_0)|}$$

$$C \equiv 5\log[r(80\%)/r(20\%)]$$

$I_0$ and $I_{180}$ represent the galaxy intensity per pixel after rotations of 0° and 180° from its original position. The value of this parameter ranges from 0 for a perfectly symmetric object to 1 for a completely asymmetric object. The percentages in the definition of the concentration parameter represent the fraction of the total intensity enclosed at the given radius. The measurements of both parameters include a correction for sky background, and were made interactively for each galaxy using an IDL-based image display and measurement program, using an adaptation of the measurement algorithm of C2000. Specifically, the sky background was taken as the mode of the counts in a rectangular annulus surrounding the target galaxy. The area of this annulus was adjusted to have several times the area of the target galaxy and placed to exclude bright nearby neighbors. This background measurement differs from that used by C2000 involving the rotation of the background area, but was found to be more practical due to the crowded nature of the HDFN and its exceptionally low background noise. Each galaxy was measured 3 - 7 times, with the final values taken as the median of the separate trials, and galaxies with > 20% variance in the trial values were excluded.

These morphological parameters were chosen for several reasons. First, C2000 and Bershady et al. (2000) find that they are closely related to, and more quantitatively constrain, traditional classification systems such as Hubble type. They are also sensitive indicators of dynamically disturbed systems (C2000; Conselice, Bershady & Gallagher 2000; see also Corbin 2000). Finally, by measuring them we can combine our data with that of C2000, who measure these parameters for 113 bright nearby galaxies that roughly span the full Hubble sequence, using the deep CCD images of Frei et al. (1996). The inclusion of these data thus effectively extends our analysis of the HDFN to the local universe.

The $(1 + z)^4$ surface brightness dimming of high-redshift galaxies will strongly affect their apparent morphologies, and so any attempt to measure them must account for this effect. Our asymmetry and concentration values are thus measured at a "metric" radius derived from the curve of growth of the galaxy brightness profile, as opposed to measuring them at a fixed isophote. This is the same method used by C2000, and is based on the parameterization first introduced by Petrosian (1976). Specifically, under the parameterization of the galaxy intensity profile $\eta(r) \equiv I(r) / <I(r)>$, where $r$ is the radial distance from the galaxy center and $<I(r)>$ is the mean intensity interior to it, C2000 measure their asymmetry parameters at radii corresponding to $\eta = 0.8$, $\eta = 0.5$, and $\eta = 0.2$. Given the much smaller angular size of the HDFN galaxies, we chose to measure their asymmetries and concentrations at only the largest radius, corresponding to $\eta = 0.2$. This turned out to introduce the largest restriction on the objects for which the asymmetry and concentration could be measured, i.e., excluding over two-thirds of the galaxies in the field because the $\eta = 0.2$ radius could not be reached. However, as will be discussed in more detail in § 2.4, this did not only eliminate galaxies with small angular sizes, and thereby introduce a selection effect. Rather, many of the brighter and larger galaxies in the field could not be measured to their $\eta = 0.2$ radii because of blending with neighbors. We are able to measure the rest-frame ultraviolet morphological parameters for a total of 105 galaxies and the rest-frame visible parameters for a total of 124 galaxies. The automatic object detection and measurement programs used by Williams et al. (1996), T99 and TWSL, namely, FOCAS and SExtractor, include many objects excluded by the $\eta = 0.2$ criterion, as they also include algorithms for object deblending. In addition to the η = 0.2 criterion, our measurements of the galaxy morphological parameters are restricted to galaxies detected at a signal-to-noise levels > 5, with a corresponding random error in the asymmetry and concentration measurements of < 20%. Due to the exceptionally low background levels in the images, most of the galaxies are detected at this signal-to-noise level.

Inspection of the galaxies in each of the individual WFPC2 and NICMOS filter images reveals that their morphologies do not vary strongly between adjacent filters. Before measuring the galaxy asymmetries and concentrations, we therefore decided to add together the images in adjacent filters to create a set of three images having higher signal-to-noise levels. Specifically, we formed the (F300W + F450W), (F606W + F814W) and (F110W + F160W) images after first aligning the images in the separate filters to within 0.1 pixel using the centroids of stars within them. The three resulting images have effective (transmission-weighted) central wavelengths of approximately 4167 Å, 6940 Å, and 1.35 μm, respectively. We measured the asymmetry and concentration of each galaxy in each of the three images whenever possible, using the catalog lists of Williams et al. (1996) and T99 to identify objects. Using the object redshift, we then determined whether these measurements covered the rest-frame ultraviolet or rest-frame visible emission of the galaxy. This was based on whether the rest-frame 4000 Å emission of the galaxy fell below or above the image central wavelength. The rest-frame UV morphology and rest-frame visible morphology of objects at $0 < z < 0.74$ are thus measured from the (F300W + F450W) and (F606W + F814W) images respectively, while for objects at $0.75 < z < 2.38$ they are measured from the (F606W + F814W) and (F110W + F160W) images. The rest-frame UV morphology of objects at $z > 2.38$ is covered by the (F110W + F160W) images, while their rest-frame visible emission is redshifted out of the available passbands. The choice of 4000 Å as a dividing point between UV and visible emission is arbitrary, but is based on the galaxy population synthesis models of Bruzual & Charlot (1993), which show that the shape of galaxy spectra below this approximate wavelength varies most strongly with galaxy age, and the associated presence of OB stars. The rest-frame emission of a given galaxy above 4000 Å is thus most likely to trace its evolved stars and underlying mass distribution.

We find good agreement between our asymmetry and concentration measurements and the qualitative appearance of the galaxies, in both the WFPC2 and NICMOS images. Examples are shown in Figure 1, which shows both the rest-frame visible images of four galaxies (extending from $z \simeq 0.1$ to $z \simeq 2.3$) and their associated brightness profiles, as traced along their major axes. The increase in the values of A(Vis) in particular matches the progression of the asymmetries evident in the brightness profiles. This provides confidence that the asymmetry parameter measures the underlying structure of the galaxies.

### 2.3 *Combined Measurements*

In Table 1 we present the measurements used in our statistical analysis, for the objects for which morphologies could be measured. The star formation rates, bolometric luminosities and fractional luminosities of additional objects can be found in TWSL. The objects are identified by their names in the Williams et al. (1996) catalog, and have been sorted in order of increasing Right Ascension. Further discussion of errors and inclusion criteria is presented in the following section. We note that the bolometric luminosity values of TWSL and the luminosity-normalized star formation rates we formed from them have been scaled to our adopted cosmology. We also include a measurement of object size, taken from Williams et al. (1996), for all objects for which the $\eta = 0.2$ radius could be reached. Specifically, we use their intensity-weighted first-moment radius measured from the (F606W + F814W) image, and convert this to kiloparsecs. While this is the most robust measure of object size available, it should be treated with the greatest caution of all the measured parameters because of the uncertainty introduced by the variance in the object asymmetries and concentrations. Finally, for each object we note which of the four galaxy spectral templates (Elliptical, Sbc, Scd and Irregular) used by Fernández-Soto et al. (1999) was found to best fit its SED.

### 2.4 *Errors, Biases, and Selection Effects*

Despite the good motivations for combining the WFPC2 and NICMOS HDFN images and Frei et al. (1996) local galaxy images for measuring morphological parameters, they form a heterogeneous data set, mainly in terms of image resolution. Measurements of them may thus be subject to systematic errors. To estimate these errors, and to assess the validity of combining the C2000 measurements with our own, we performed the following tests. First, since our algorithm for the measurement of asymmetry and concentration differs slightly from that of C2000 in terms of the image background, we independently measured the asymmetries and concentrations of a random subset of galaxies in the Frei et al. (1996) images, and found agreement between our values and those of C2000 to within ~ 20%. As in the case of our measurements, there is also good agreement between the values of the asymmetry and concentration parameters measured by C2000 and Bershady et al. (2000) and the qualitative appearance of the galaxies, which further supports the combination of those measurements with our own. Concerning the WFPC2 and NICMOS images, since they differ in angular resolution by a factor of approximately two, we tested for any systematic effect caused by this difference by re-sampling the WFPC2 images to the resolution of the NICMOS images, and comparing the corresponding asymmetry and concentration measurements for a sample of approximately 15 randomly selected galaxies. We find no strong systematic differences in these measurements, although we confirm the result of C2000 and Wu, Faber & Lauer (1997) that the asymmetry values of elliptical galaxies tend to *increase* slightly as image resolution is lowered. The mean difference between the morphological parameters measured from the original and re-sampled WFPC2 images is approximately 20%. This indicates a source of uncertainty in the combined set of WFPC2 and NICMOS morphological measurements in addition to the relatively low errors introduced by Poisson noise in the galaxies and image background (§ 2.2). We thus estimate total random errors for the combined set of C2000, WPFC2 and NICMOS asymmetry and concentration parameters to be relatively large, in the range ~ 10% - 30%. A detailed analysis (including Monte Carlo simulations) of the possible errors in the galaxy bolometric luminosities and star formation rates is presented by TWSL. In particular, they address the issue of the degeneracy in SEDs produced by dust extinction versus the age of the stellar population and how this affects their estimates of star formation rates and bolometric luminosities.

The Frei et al. (1996) galaxy sample is not complete in any sense, and so the combination of the morphological measurements of C2000 and Bershady et al. (2000) with those from the HDFN galaxies must be carefully considered. In particular, the Frei et al. sample contains only a few dwarf irregular galaxies. However, independent of the issue of how well the local galaxy population is in fact characterized, it can be said that the Frei et al. sample covers the full range of Hubble types without a strong bias towards any one part of the sequence (see Frei et al. and C2000). This sample also contains several strongly asymmetric galaxies such as Arp 18 (NGC 4088) and NGC 4731, which may have recently undergone mergers. The Frei et al. sample is therefore not biased toward symmetric galaxies, which is important insofar as such a bias could create a false correlation between redshift and asymmetry when measurements of these galaxies are combined with those of galaxies in the HDFN. We thus proceed under the assumption that the Frei et al. sample is, to first order, representative of the local galaxy population. The lack of low-luminosity irregular galaxies in this sample may in fact be beneficial for the present comparison, since this will tend to compensate for any selection against such objects at high redshift, although, as we discuss below, we find no strong evidence of incompleteness in the HDFN sample.

As noted above, the exclusion of objects in the HDFN for which the $\eta = 0.2$ radius could not be reached would seem to introduce a bias against very high redshift, intrinsically compact and low surface brightness galaxies. However, in practice an equally important factor in whether this radius could be reached is to what degree the galaxy is isolated. The $\eta = 0.2$ radius reaches close to the outer edges of the disks of normal spiral galaxies (see C2000). Consequently, many of the apparently larger galaxies in the field could not be measured to $\eta = 0.2$ because of blending with nearby objects. This also affected smaller and fainter objects. The range of bolometric luminosities over which we are able to measure asymmetries and concentrations still extends over approximately 3.3 dex, which argues against a strong luminosity bias in the sample. To more quantitatively assess this, we performed the two-sided Kolmogorov-Smirnov test on the bolometric luminosity distributions of objects for which A(Vis) could and could not be measured. We find that these distributions differ at only the 82% confidence level. As the total TWSL sample is ~ 20% - 30% complete compared to the local luminosity function to $z = 2$, we thus assume a similar completeness in the sub-sample for which A(Vis) is measured. The remaining incompleteness is likely due to excluding low surface brightness and/or low luminosity irregular galaxies, as well as galaxies with high levels of internal extinction which consequently have reduced rest-frame UV emission (see TWSL). As presented in the next section, we also find no correlation between redshift and galaxy size among the galaxies for which we are able to measure the morphological parameters, which argues that we have not selected against small galaxies at high redshift.

Finally, an analysis by Storrie-Lombardi, Thompson & Weymann (1999) indicates that ~ 95% of the most compact objects (those having areas $< 0.2$ arcsec$^2$) in the T99 portion of the HDFN are likely to be at $z > 2$, which is close to the limit at which our (F110W + F160W) NICMOS images still cover the rest-frame visible continua of these galaxies. Their exclusion thus does not seriously harm our effort to trace the evolution of the rest-frame visible morphologies of these galaxies, which as discussed previously should be a better indicator of their underlying mass distribution. The HDFN galaxies also show peaks in their redshift distribution, mainly at $z \cong 0.5$ and $z \cong 1$ (Cohen et al. 1996), which is likely the effect of large-scale structure within the pencil beam of the field. The effect of these peaks on the present analysis is unclear, but any bias may again be compensated by the fact that our morphological measurements exclude some of the apparently larger and more grouped galaxies due to the overlap of their brightness profiles.

## 2. ANALYSIS AND RESULTS

We ran the Spearman rank correlation test on all pairs of measured parameters listed in Table 1. For the comparisons involving A(Vis) and C(Vis), we include the asymmetry and concentration parameters measured by C2000 and Bershady et al. (2000) at the $\eta = 0.2$ radii of the $R$ and $r$ band images of Frei et al. (1996) for 113 nearby galaxies. The resulting correlation matrix is given in Table 2, which lists the Spearman rank test correlation coefficient. Values of this coefficient $> 0.4$ indicate a correlation at $> 99\%$ confidence, while values $> 0.5$ indicate a correlation at $> 99.99\%$ confidence. The number of objects involved in the individual tests varies from a maximum of 276 to a minimum of 33. The principal components analysis of this correlation matrix is presented in Table 3, which lists the coefficients of the first four eigenvectors. We consider first the results of the individual correlation tests, given the relatively few cases where a correlation is indicated at high ($> 99.99\%$) confidence.

Comparisons of the A(Vis), A(UV), C(Vis) and C(UV) measurements offer additional diagnostic tests of these parameters. These comparisons are shown in Figure 2, which shows a weak correlation in the case of the A(Vis), A(UV) comparison and a stronger correlation between C(Vis) and C(UV). This is consistent with the qualitative findings of Bunker (1999) and Dickinson (1999) that there is a general agreement between the morphologies of HDFN galaxies in the rest-frame UV and visible. Importantly, the C(Vis) values are larger on average than the corresponding C(UV) values. This is consistent with the expectation that spiral galaxies will appear less concentrated in the rest-frame UV, as the emission in that regime will be dominated by star-forming regions in their disks. Similarly, the A(UV) measurements are skewed to slightly higher values than the A(Vis) measurements, also consistent with having the galaxies' UV emission dominated by irregularly distributed regions of star formation (e.g., the galaxy HDFN 4-378; see Bunker 1999). The results of Figure 2 thus provide additional confidence that the chosen wavelength dividing point between UV and visible adequately discriminates between the early type and late type stars in these galaxies.

Several of the correlations with redshift are due to unavoidable selection effects. Specifically, the apparent correlations between redshift and bolometric luminosity, star formation rate, and mid-infrared fractional luminosity are likely due, as noted above, to the non-detection of dwarf galaxies, low surface brightness galaxies, and heavily extincted galaxies at $z > 1$. However, the strongest correlation revealed within the sample, between bolometric luminosity and star formation rate, is intrinsically meaningful. This comparison is shown in Figure 3. What is particularly noteworthy in this comparison is the clear deviation of the galaxies whose SEDs indicate that they are dominated by evolved stellar populations, as determined by TWSL. Not surprisingly, inspection of their images reveals that these galaxies are the most elliptical in appearance, and have the reddest colors. We discuss the possible significance of this result in the next section. The strength of this correlation clearly reduces the use of the luminosity-normalized star formation rate parameter, as this ratio is thus nearly constant.

Only two other pairs of variables are correlated at > 99.99% confidence. The first is bolometric luminosity and galaxy radius. This is not unexpected, and can be better constrained from samples of nearby galaxies. Given the related uncertainties inherent in the radius measurements (§ 2.2), this result will not be considered further. The second correlation is between redshift and A(Vis). The significance of this correlation is dependent on the inclusion of the C2000 asymmetry values: if this comparison is restricted to only the HDFN galaxies, the Spearman rank test coefficient falls to 0.250. Yet as discussed above, we believe that this inclusion is valid, based in part on our ability to reproduce the C2000 values for the same data, and the wide range of Hubble types and asymmetries covered by this sample. However, given the large uncertainties associated with the asymmetry measurements for the combined sample (§ 2.4), as well as the uncertainties in the associated photometric redshifts, the putative correlation should only be regarded as suggestive. However, it can be noted that any systematic bias in the asymmetry measurements of the highest redshift ($z \sim 2$) galaxies in the sample would be that their asymmetry values are *underestimated*, as this would be the result if the NICMOS images do not adequately resolve them (§ 2.2). Higher asymmetry values for these galaxies would strengthen the suggested correlation.

The comparison of redshift and A(Vis) is shown in Figure 4. In Figure 4b we show the mean values of these data after binning them in redshift intervals of approximately 0.5 (the actual size of the intervals was selected to maintain sub-samples of approximately equal numbers of objects, to within the irregularity in the redshift distribution). These mean values show a monotonic increase that is most clear when redshift is converted to look-back time. This is shown in Figure 5, where it can be seen that the mean asymmetry values increase by a factor of approximately three from the present epoch to $z \cong 2$. Two points are worth noting. First, the scatter in the comparison of the individual measurements (Figure 4a) is obviously large, and it is possible to find galaxies as symmetric and asymmetric as the those in the nearby galaxy sample at any redshift up to $z \cong 2$. However, there are no galaxies in the HDFN that match the lower asymmetry range of the local sample. In particular, all the elliptical galaxies in the HDFN have asymmetry values slightly above their local counterparts. This may, however, be an artifact of their higher redshifts. Specifically, as noted previously, our simulations, as well as those of C2000 and Wu, Faber & Lauer (1997) show that while the initial decrease in the angular size of disk galaxies with redshift tends to lower their asymmetry values, for elliptical galaxies the opposite (and counter-intuitive) result is obtained, i.e., their asymmetries *increase* slightly as resolution is lowered. We return to further discussion of this result in the next section.

We lastly consider the results of the principal components analysis (Table 3). The lack of a large number of correlations between the measured parameters, and the existence of the observationally-biased correlations with redshift makes these results of limited use. It is however interesting to note that the first eigenvector is dominated by the anticorrelation of the radius and concentration parameters, although the individual correlations are not very strong (Table 2). The mid-infrared fractional luminosity and luminosity-normalized star formation rate also appear to be involved in this relation. The second and third eigenvectors are dominated by the individual correlations between bolometric luminosity and star formation rate, and between redshift and A(Vis). The fourth eigenvector is most strongly related to radius and normalized star formation rate.

## 4. DISCUSSION

The interpretation of the one strong correlation we find, between bolometric luminosity and star formation rate, must take into account that both of these measurements are *indirect*, and to first order represent quantities scaled from the fitted galaxy spectral templates. It is thus not surprising that they are correlated at some level. As discussed previously, the sample could also be slightly biased against actively star forming but intrinsically faint galaxies such as Magellenic-type irregulars, which would weaken the correlation. However, to the extent that both quantities are accurately represented, this correlation shows the dominance of OB stars and active star formation on the value of bolometric luminosity. The deviation of luminous early-type galaxies from the main locus of points evident in Figure 3 thus suggests a more advanced evolutionary state for them than their late-type counterparts. This is consistent with the "downsizing" picture first introduced by Cowie et al. (1996; see also Guzmán et al. 1997 and Balland, Silk & Schaeffer 1998) in which more massive galaxies are the first to form. However, these results do not necessarily imply a monolithic collapse, as opposed to hierarchical, formation process for early-type galaxies. As noted above, the elliptical galaxies in the HDFN are more asymmetric than their local counterparts, which may or may not be an artifact of their reduced angular size. If this effect is real, it suggests that these elliptical galaxies, which occur in the HDFN mainly at $z < 1$ (see Fig. 4a) are not as dynamically relaxed as elliptical galaxies in the local universe. Menanteau et al. (1999) also find evidence of lingering star formation among faint elliptical galaxies identified in archival WFPC2 images, which is inconsistent with a monolithic collapse at high redshift, and Corbin et al. (2000a) identify several galaxies in the NICMOS parallel fields which appear to be elliptical galaxies in the process of merging. A hierarchical formation process may thus apply to both disk and spheroidal galaxies, with the latter simply beginning the process earlier than the former, and also more efficiently converting their gas to stars (cf. Balland et al. 1998).

We proceed with the interpretation of the suggested trend between redshift and rest-frame visible asymmetry under the assumption that this trend is, while weak, a real effect involving evolution in galaxy structure. If this result is spurious, then it indicates that either previous claims of galaxy morphological evolution (§ 1) are in error due to the lack of morphological *K*-corrections, or else that the chosen asymmetry parameter fails to quantify this evolution. However, in support of the view that this relation is real, we note the results of Brinchmann & Ellis (2000), who have approached this problem by making estimates of the masses of galaxies out to $z \sim 1$, and find a strong evolution in these masses for galaxies classified as peculiar or interacting. Indeed, a direct comparison of redshift and galaxy masses provides the most fundamental test of hierarchical formation models, and will be pursued with the present data in later studies. A drawback for such comparisons however is that the mass estimates are model dependent, and subject to the array of errors inherent in the galaxy flux measurements and SED-template fittings (see TWSL). The asymmetry parameter offers a more direct measurement, while not directly constraining a physical quantity. Measurement of this asymmetry parameter for the *N*-body simulations of galaxy formation and mergers (e.g. Walker, Mihos & Hernquist 1996; Contardo, Steinmetz & Fritze-von Alvensleben 1998) would be very useful as a means of matching them to these observational results.

The salient feature of the look-back time / asymmetry relation (Figure 5) is its continuous and roughly linear form, extending through the present epoch as defined by the nearby galaxy sample. This strongly suggests that galaxy morphological evolution is a gradual process that is continuing through the present epoch, as opposed to a relatively short (~ 1-4 Gyr) formation/relaxation epoch above $z \sim 1$ followed by little or no structural changes. Several lines of evidence suggest that the "minor" mergers of large disk galaxies with smaller companions dominate this process. The first is the qualitative similarity of the $N$-body simulations of such mergers by Walker et al. (1996) to the appearance of many HDFN galaxies. These simulations yield asymmetric galaxy morphologies similar to those observed (e.g. HDFN 3-430.1 and HDFN 4-660.0; Fig. 1a) within a period ~ 0.5 Gyr before the completion of the merger. A large number of the galaxies in the present sample, e.g. HDFN 4-558.0 (Fig. 1a) show evidence of being in the early stage of such mergers, and have small, marginally resolved companions. Walker et al. (1996) find that the effect of such mergers is to effectively transform late-type spiral galaxies into early-type spiral galaxies by enlarging the galaxy bulges upon their completion. This would account for the increase in the fraction of late-type spiral galaxies with redshift noted by both Driver et al. (1998) and Im et al. (1999), and the result of Brichmann & Ellis (2000) that morphological evolution is coupled to changes in galaxy mass. Such a model is also consistent with the evolution in the incidence of galaxy pairs (see the references in § 1), modulo the uncertainties in the relative masses of the pair members. Finally, there is ample evidence from nearby spiral galaxies that minor mergers are affecting, and will continue to affect, galaxy morphologies. First, Zaritsky et al. (1997) have found that ~ 75% of isolated spiral galaxies have at least one small nearby companion, and such companions are likely to merge with the host within the next few Gyr. Second, Zaritsky & Rix (1997) and Rudnick & Rix (1998) find that the incidence of nearby spiral galaxies in which a minor merger may have recently (within < 1 Gyr) occurred is ~ 20% - 30%, as judged from the asymmetry of their disks. The continuous decrease in the mean asymmetry values seen in Figure 5 would then indicate a concurrent decrease in the galaxy merger rate with time, a conclusion also reached by Carlberg et al. (2000) and Le Fevre et al. (2000) on the basis of the redshift-dependence of the number of galaxy pairs.

This is not to say that major mergers of larger disk and spheroidal galaxies play no role in the observed morphological evolution. Indeed, TWSL find that two HDFN galaxies (4-186.0 and 4-307.0) have luminosities that qualify them as Ultraluminous Infrared Galaxies (ULRIGs), which at low redshift show clear evidence of being major mergers (see Surace, Sanders & Evans 2000 and references therein). ULIRGs in the range $1 < z < 2$ may comprise a significant fraction of the population of Extremely Red Objects, which itself appears to be a significant fraction of the galaxy population in this redshift range (see Corbin et al. 2000b and references therein). However, the continuity between the asymmetry values of local galaxies not undergoing major mergers with their high-redshift counterparts (Fig. 5), along with the simple fact that most local galaxies do not appear to be the products of major mergers, suggests that minor mergers are the dominant source of the observed evolution.

Finally, we comment on the absence of correlations that we suspected, *ab initio*, of being present. There is no strong correlation between redshift and C(Vis) (Table 2), which could be expected under the interpretation that the trend between redshift and asymmetry is driven by merging. That is, if as the models of Walker et al. (1996) show, minor mergers increase the bulge sizes of spiral galaxies, then before such mergers are complete such galaxies would appear less concentrated. The lack of such a correlation could be because of the choice of the outer and inner radii used in the concentration parameter, and/or could indicate a more complex relationship between morphology and minor mergers than the Walker et al. models imply. The lack of any evidence of correlation between asymmetry and star formation rate is also somewhat surprising. Le Fevre et al. (2000) have claimed that mergers boost the star formation rates of galaxies at intermediate redshifts by a factor of roughly two, as derived from [O II] line equivalent widths. Among nearby galaxies, Rudnick, Rix & Kennicutt (2000) find a similar (but smaller) increase in the star formation rates of galaxies that appear to have recently (within the last ~ 1 Gyr) undergone minor mergers. A possible explanation for the lack of such a correlation among our galaxies is the evidence that the star formation of high-redshift galaxies proceeds episodically. Specifically, Sawicki & Yee (1998; see also Somerville, Primack & Faber 2000) find that the $z > 2$

Lyman break galaxies in the HDFN are dominated by very young (~ 25 Myr) stellar populations. Such an age is short relative to the time scale over which minor mergers produce strong asymmetries (~ 500 Myr; see Walker et al. 1996). Thus if merging produces a series of bursts of such short duration followed by longer periods of quiescence, it would effectively remove any correlation between star formation rate and asymmetry. This could also contribute to the lack of a correlation between bolometric luminosity and luminosity-normalized star formation rate similar to that found between galaxy mass and mass-normalized star formation rate (Guzmán et al. 1997; Brinchmann & Ellis 2000). That is, given the evidence that short-lived early type stars dominate the bolometric luminosities of these galaxies (Fig. 3), episodic star formation would serve to obscure any mass / luminosity correlation, and consequently any correlations involving the luminosity-normalized star formation rate. Mass estimates of more galaxies above $z \sim 1$ will be required to test to what epochs the correlations found by Guzmán et al. (1997) and Brinchmann & Ellis (2000) extend. Alternatively, or perhaps additionally, the relation between mergers and star formation inferred among local galaxies may not apply at earlier epochs, particularly if the stellar and neutral hydrogen mass distributions of galaxies at such epochs are significantly different.

We thank the referee, Harry Ferguson, for comments and suggestions that improved this paper. We also thank Chris Conselice for helpful discussions of asymmetry and concentration parameters, and for providing us with copies of his scripts for measuring them, and Mark Dickinson for help with his NICMOS HDFN images. Helpful discussions of the results were provided by Rob Kennicutt, Luc Simard, Matthias Steinmetz and Katherine Wu. This work was supported by NASA grant NAG5-3042 to the University of Arizona.

TABLE 1

MEASURED PARAMETERS FOR HDFN GALAXIES[1]

| ID | z | $r_1$ | A(UV) | C(UV) | A(Vis) | C(Vis) | SFR | log $L_{bol}$ | log(SFR/L) | frac | Sp. Typ |
|---|---|---|---|---|---|---|---|---|---|---|---|
| 4-916.0 | 0.16 | 1.0 | ----- | ----- | 0.124 | 3.75 | ----- | ----- | ------ | ----- | 3 |
| 4-950.0 | 0.609 | 3.1 | 0.529 | 2.19 | 0.408 | 2.62 | ----- | ----- | ------ | ----- | 3 |
| 4-942.0 | 1.00 | 1.4 | 0.159 | 3.91 | 0.311 | 4.47 | ----- | ----- | ----- | ----- | 2 |
| 4-823.0 | 0.64 | 1.4 | ----- | ----- | 0.152 | 3.69 | ----- | ----- | ----- | ----- | 3 |
| 4-801.0 | 0.92 | 1.5 | 0.467 | 3.70 | 0.482 | 4.37 | ----- | ----- | ----- | ----- | 4 |
| 4-928.0 | 1.015 | 2.3 | 0.166 | 3.87 | ------ | ----- | ----- | ----- | ----- | ----- | 2 |
| 4-888.0 | 1.01 | 1.1 | 0.505 | 2.85 | ----- | ----- | ----- | ----- | ----- | ----- | 4 |
| 4-878.0 | 0.00: | ----- | 0.573 | 3.75 | 0.379 | 3.95 | ----- | ----- | ----- | ----- | 2 |
| 4-822.0 | 0.16 | 0.8 | ----- | ----- | 0.367 | 3.13 | 0.615 | 10.15 | -10.36 | 0.95 | 3 |
| 4-948.0 | 0.585 | 6.7 | ----- | ----- | 0.155 | 3.42 | ----- | ----- | ----- | ----- | 2 |
| 4-767.0 | 0.72 | 1.6 | ----- | ----- | 0.135 | 4.16 | 0.019 | 10.97 | -12.70 | -0.02 | 1 |
| 4-794.0 | 0.80 | 1.2 | 0.770 | 2.86 | ----- | ----- | 0.432 | 10.11 | -10.47 | 0.47 | 4 |
| 4-976.1 | 0.089 | 1.8 | ----- | ----- | 0.297 | 2.82 | ----- | ----- | ----- | ----- | 4 |
| 4-661.0 | 0.52 | 1.0 | ----- | ----- | 0.170 | 3.73 | ----- | ----- | ----- | ----- | 3 |
| 4-639.1 | 0.00: | ----- | 0.684 | 3.86 | 0.473 | 4.06 | ----- | ----- | ----- | ----- | 4 |
| 4-795.0 | 0.40 | 2.6 | ----- | ----- | 0.129 | 2.65 | 0.460 | 10.90 | -11.24 | 0.23 | 3 |
| 4-665.0 | 1.44 | 3.1 | ----- | ----- | 0.102 | 4.75 | 3.602 | 11.01 | -10.45 | 0.60 | 4 |
| 4-769.0 | 0.96 | 1.6 | 0.285 | 2.73 | 0.540 | 3.45 | 1.362 | 10.53 | -10.40 | 0.55 | 4 |
| 4-671.0 | 0.96 | 1.2 | ----- | ----- | 0.339 | 4.22 | 1.556 | 10.63 | -10.44 | 0.75 | 4 |
| 4-690.0 | 1.12 | 1.1 | ----- | ----- | 0.148 | 4.54 | 2.459 | 10.78 | -10.39 | 0.88 | 4 |
| 4-602.0 | 2.04 | 1.1 | 0.162 | 4.06 | ----- | ----- | ----- | ----- | ----- | ----- | 4 |
| 4-619.0 | 2.96 | 0.8 | 0.762 | 3.84 | ----- | ----- | 1.219 | 10.69 | -10.61 | 0.42 | 4 |
| 4-636.0 | 0.64 | 0.9 | ----- | ----- | ----- | ----- | 1.675 | 10.66 | -10.44 | 0.88 | 4 |
| 4-725.0 | 1.84 | 1.6 | ----- | ----- | 0.288 | 3.59 | 0.661 | 10.38 | -10.56 | 0.00 | 4 |
| 4-581.0 | 1.92 | 1.3 | ----- | ----- | ----- | ----- | 6.356 | 11.22 | -10.42 | 0.50 | 4 |
| 4-697.0 | 2.48 | 1.1 | 0.166 | 4.18 | ----- | ----- | 1.818 | 10.78 | -10.52 | 0.12 | 4 |
| 4-656.0 | 0.56 | 3.5 | 0.499 | 2.14 | 0.196 | 2.12 | 67.57 | 12.22 | -10.39 | 0.91 | 3 |
| 4-660.0 | 2.32 | 1.4 | 0.613 | 3.55 | 0.801 | 3.73 | 3.257 | 10.93 | -10.42 | 0.13 | 4 |
| 4-554.1 | 0.84 | 1.5 | 0.271 | 3.6 | ----- | ----- | ----- | ----- | ----- | ----- | 4 |
| 4-493.0 | 0.847 | 2.4 | ----- | ----- | 0.221 | 4.98 | ----- | ----- | ----- | ----- | 1 |
| 4-775.0 | 1.12 | 3.4 | 0.309 | 3.04 | 0.204 | 3.67 | ----- | ----- | ----- | ----- | 2 |
| 4-590.0 | 2.08 | 1.3 | ----- | ----- | 0.144 | 4.24 | 21.32 | 11.74 | -10.41 | 0.79 | 4 |
| 4-727.0 | 1.242 | 1.7 | 0.140 | 3.21 | 0.337 | 4.20 | ----- | ----- | ----- | ----- | 4 |
| 4-565.0 | 0.56 | 2.0 | 0.286 | 3.11 | 0.207 | 3.06 | 21.52 | 11.72 | -10.39 | 0.86 | 3 |
| 4-572.0 | 0.48 | 1.7 | ----- | ----- | ----- | ----- | 1.051 | 10.34 | -10.32 | 0.75 | 4 |
| 4-479.0 | 1.12 | 1.1 | 0.393 | 3.94 | ----- | ----- | 1.512 | 10.61 | -10.43 | 0.35 | 4 |
| 4-744.0 | 0.764 | 3.3 | ----- | ----- | 0.178 | 4.68 | ----- | ----- | ----- | ----- | 1 |
| 4-439.1 | 4.32 | 0.9 | 0.180 | 3.11 | ----- | ----- | ----- | ----- | ----- | ----- | 4 |
| 4-402.3 | 0.557 | 2.9 | 0.344 | 2.22 | 0.176 | 2.41 | ----- | ----- | ----- | ----- | 2 |
| 2-82.1 | 2.267 | 1.8 | 0.756 | 3.91 | ----- | ----- | ----- | ----- | ----- | ----- | 4 |
| 1-54.2 | 2.929 | 2.1 | ----- | ----- | ----- | ----- | ----- | ----- | ----- | ----- | 4 |
| 4-500.0 | 1.28 | 1.3 | ----- | ----- | 0.677 | 3.21 | 0.055 | 9.51 | -10.77 | 0.00 | 4 |
| 4-430.0 | 0.72 | 2.9 | 0.748 | 2.98 | 0.488 | 2.20 | 0.870 | 10.46 | -10.52 | 0.26 | 4 |
| 4-402.0 | 0.558 | 10.7 | 0.668 | 3.21 | 0.419 | 2.89 | ----- | ----- | ----- | ----- | 4 |
| 4-752.1 | 1.013 | 4.5 | 0.119 | 3.58 | ----- | ----- | ----- | ----- | ----- | ----- | 1 |
| 4-627.0 | 0.16 | 0.6 | ----- | ----- | ----- | ----- | 0.170 | 9.59 | -10.36 | 0.95 | 4 |
| 4-505.1 | 1.28 | 3.2 | ----- | ----- | 0.122 | 3.15 | 18.83 | 11.60 | -10.32 | 0.87 | 2 |
| 4-527.0 | 0.96 | 1.1 | ----- | ----- | 0.505 | 4.08 | 0.658 | 10.16 | -10.32 | 0.88 | 3 |
| 4-445.0 | 1.84 | 2.1 | 0.34 | 3.40 | 0.239 | 4.25 | 80.45 | 12.29 | -10.38 | 0.75 | 4 |
| 4-579.0 | 0.48 | 1.1 | 0.299 | 3.96 | 0.204 | 3.53 | 1.103 | 10.52 | -10.48 | 0.56 | 4 |
| 4-603.0 | 2.56 | 1.4 | 0.694 | 3.91 | ----- | ----- | 2.072 | 10.73 | -10.42 | 0.12 | 4 |
| 4-558.0 | 0.48 | 2.8 | ----- | ----- | 0.311 | 2.67 | 5.404 | 11.05 | -10.32 | 0.84 | 3 |
| 4-509.0 | 1.12 | 1.2 | ----- | ----- | 0.599 | 4.37 | 0.20 | 9.59 | -11.29 | 0.00 | --- |
| 4-351.0 | 2.48 | 0.9 | ----- | ----- | ----- | ----- | 0.626 | 10.37 | -10.57 | 0.11 | 4 |
| 4-378.0 | 1.12 | 3.8 | ----- | ----- | 0.370 | 2.72 | 0.499 | 10.46 | -10.77 | 0.07 | 4 |
| 4-596.0 | 0.48 | 1.0 | ----- | ----- | 0.837 | 3.25 | 0.126 | 9.58 | -10.48 | 0.23 | 4 |
| 4-316.0 | 1.76 | 4.0 | 0.484 | 2.97 | ----- | ----- | ----- | ----- | ----- | ----- | 4 |
| 4-571.0 | 0.72 | 2.1 | 0.546 | 2.60 | 0.370 | 2.64 | 0.503 | 10.25 | -10.55 | 0.32 | 4 |
| 4-543.0 | 1.28 | 1.8 | 0.983 | 3.26 | 0.817 | 3.93 | 5.356 | 11.08 | -10.35 | 0.60 | 4 |
| 4-502.0 | 2.00 | 2.0 | ----- | ----- | 0.477 | 3.10 | 7.019 | 11.14 | -10.29 | 0.50 | 4 |
| 4-593.0 | 0.40 | 0.0 | ----- | ----- | 0.220 | 2.83 | 0.781 | 10.33 | -10.44 | 0.92 | 3 |
| 4-407.0 | 0.56 | 1.5 | ----- | ----- | 0.533 | 3.66 | 2.548 | 10.77 | -10.36 | 0.88 | 4 |
| 4-498.0 | 1.92 | 1.1 | 0.646 | 3.4 | 0.608 | 4.45 | 1.694 | 10.74 | -10.51 | 0.46 | 4 |
| 4-395.0 | 0.72 | 1.0 | ----- | ----- | 0.58 | 3.04 | 1.285 | 10.52 | -10.41 | 0.60 | 4 |
| 4-555.1 | 3.12 | 3.8 | 0.19 | 4.46 | ----- | ----- | 15.28 | 11.81 | -10.63 | 0.00 | 4 |
| 4-368.0 | 1.92 | 1.8 | ----- | ----- | 0.386 | 3.85 | 17.11 | 11.61 | -10.38 | 0.73 | 3 |

| | | | | | | | | | | |
|---|---|---|---|---|---|---|---|---|---|---|
| 4-557.0 | 2.16 | 1.6 | 0.273 | 3.60 | 0.503 | 4.43 | 1.365 | 10.97 | -10.84 | 0.00 | 4 |
| 4-344.0 | 1.20 | 1.0 | ----- | ----- | 0.590 | 3.85 | 1.659 | 10.61 | -10.39 | 0.58 | 4 |
| 4-345.0 | 1.36 | 1.0 | ----- | ----- | 0.310 | 4.65 | 1.870 | 10.73 | -10.46 | 0.56 | 4 |
| 4-389.0 | 2.72 | 1.1 | ----- | ----- | ----- | ----- | 4.874 | 11.06 | -10.37 | 0.42 | 4 |
| 4-516.0 | 1.04 | 1.6 | 0.532 | 4.19 | 0.239 | 4.52 | 1.329 | 11.00 | -10.88 | 0.31 | 3 |
| 4-307.0 | 1.60 | 2.1 | ----- | ----- | 0.267 | 4.36 | 534.70 | 131.13 | -10.40 | 0.97 | 1 |
| 4-563.0 | 4.40 | 1.4 | 0.354 | 4.57 | ----- | ----- | 2.847 | 11.40 | -10.95 | 0.17 | 2 |
| 4-497.0 | 2.16 | 1.0 | 0.340 | 3.81 | ----- | ----- | ----- | ----- | ----- | ----- | 4 |
| 2-239.0 | 2.427 | 4.3 | ----- | ----- | ----- | ----- | ----- | ----- | ----- | ----- | 4 |
| 4-473.0 | 5.52 | 0.6 | 0.496 | 4.35 | ----- | ----- | 4.507 | 11.00 | -10.35 | 0.00 | 4 |
| 4-305.0 | 0.96 | 1.6 | 0.285 | 3.31 | 0.683 | 3.89 | 4.937 | 11.06 | -10.37 | 0.88 | 3 |
| 4-460.0 | 0.56 | 2.0 | ----- | ----- | 0.252 | 2.80 | 7.682 | 11.23 | -10.35 | 0.86 | 4 |
| 4-254.0 | 0.60 | 2.2 | 0.333 | 3.20 | ----- | ----- | ----- | ----- | ----- | ----- | 1 |
| 4-522.0 | 1.12 | 1.4 | ----- | ----- | ----- | ----- | 7.191 | 11.23 | -10.38 | 0.79 | 4 |
| 4-550.0 | 1.04 | 5.5 | 0.817 | 2.02 | 0.209 | 2.77 | 1.250 | 11.43 | -11.33 | 0.17 | 3 |
| 4-448.0 | 0.48 | 2.2 | ----- | ----- | 0.451 | 2.82 | 3.561 | 10.82 | -10.27 | 0.92 | 3 |
| 4-300.0 | 0.56 | 1.4 | 0.926 | 3.19 | ----- | ----- | 1.424 | 10.60 | -10.45 | 0.73 | 3 |
| 2-251.0 | 0.96 | 3.0 | 0.184 | 4.75 | 0.105 | 4.96 | ----- | ----- | ----- | ----- | 2 |
| 4-327.0 | 1.84 | 1.3 | 0.915 | 2.90 | 0.552 | 3.96 | 9.414 | 11.35 | -10.38 | 0.79 | 4 |
| 4-303.0 | 1.92 | 0.9 | ----- | ----- | 0.605 | 4.53 | 1.606 | 10.65 | -10.44 | 0.69 | ----- |
| 2-270.0 | 0.16 | 1.0 | 0.225 | 3.94 | 0.200 | 3.97 | ----- | ----- | ----- | ----- | 4 |
| 4-471.0 | 0.32 | 1.6 | ----- | ----- | ----- | ----- | 0.009 | 10.48 | -12.53 | 0.00 | 1 |
| 4-161.0 | 0.92 | 1.7 | 0.103 | 3.25 | ----- | ----- | ----- | ----- | ----- | ----- | 4 |
| 4-488.0 | 2.48 | 1.1 | 0.976 | 5.18 | ----- | ----- | 3.169 | 10.92 | -10.42 | 0.14 | 4 |
| 4-416.0 | 0.454 | 1.4 | ----- | ----- | 0.209 | 3.48 | 3.081 | 10.91 | -10.42 | 0.77 | 4 |
| 4-434.0 | 0.16 | 0.6 | 0.909 | 4.04 | ----- | ----- | 0.227 | 9.77 | -10.42 | 0.88 | 4 |
| 4-350.0 | 1.20 | 1.1 | 0.127 | 3.94 | 0.20 | 4.50 | 0.079 | 9.85 | -10.95 | 0.00 | 4 |
| 4-475.0 | 3.68 | 0.8 | ----- | ----- | ----- | ----- | 3.183 | 10.95 | -10.45 | 0.12 | 4 |
| 4-442.0 | 0.64 | 1.4 | ----- | ----- | 0.383 | 3.01 | 0.232 | 10.02 | -10.65 | 0.38 | 4 |
| 4-241.0 | 0.48 | 9.4 | ----- | ----- | 0.247 | 4.33 | ----- | ----- | ----- | ----- | 4 |
| 4-474.0 | 1.059 | 6.1 | ----- | ----- | 0.493 | 2.68 | ----- | ----- | ----- | ----- | 4 |
| 4-382.0 | 0.08 | 0.5 | 0.121 | 3.10 | 0.149 | 3.01 | 0.269 | 9.78 | -10.35 | 0.95 | 3 |
| 4-289.0 | 2.88 | 1.4 | ----- | ----- | ----- | ----- | 6.719 | 11.23 | -10.40 | 0.00 | 4 |
| 4-241.1 | 0.321 | 2.0 | 0.494 | 4.03 | ----- | ----- | ----- | ----- | ----- | ----- | 4 |
| 4-415.0 | 0.48 | 1.3 | 0.348 | 3.93 | 0.742 | 2.60 | 0.460 | 10.08 | -10.42 | 0.56 | 4 |

| | | | | | | | | | | |
|---|---|---|---|---|---|---|---|---|---|---|
| 4-332.0 | 0.48 | 1.4 | ----- | ----- | 0.610 | 3.34 | 1.440 | 10.57 | -10.41 | 0.58 | 4 |
| 2-353.0 | 0.609 | 1.5 | ----- | ----- | 0.189 | 3.07 | ----- | ----- | ----- | ----- | 2 |
| 2-201.0 | 1.16 | 1.6 | 0.43 | 4.86 | ----- | ----- | ----- | ----- | ----- | ----- | 4 |
| 4-385.0 | 0.16 | 0.6 | ----- | ----- | ----- | ----- | 0.167 | 9.63 | -10.41 | 0.88 | 4 |
| 4-232.0 | 0.40 | 2.6 | ----- | ----- | ----- | ----- | 7.773 | 11.23 | -10.34 | 0.86 | 4 |
| 4-298.0 | 2.64 | 1.1 | ----- | ----- | ----- | ----- | 1.804 | 10.67 | -10.42 | 0.27 | 4 |
| 4-319.0 | 0.72 | 1.3 | ----- | ----- | 0.683 | 2.69 | 1.988 | 10.68 | -10.38 | 0.88 | 4 |
| 4-346.0 | 0.72 | 2.0 | ----- | ----- | ----- | ----- | 0.396 | 10.19 | -10.59 | 0.18 | 4 |
| 4-89.0 | 0.681 | 2.4 | ----- | ----- | 0.567 | 3.43 | ----- | ----- | ----- | ----- | 4 |
| 4-173.0 | 0.929 | 2.0 | 0.171 | 2.70 | ----- | ----- | ----- | ----- | ----- | ----- | 3 |
| 4-212.0 | 1.12 | 1.7 | ----- | ----- | 0.475 | 4.08 | 0.963 | 10.42 | -10.44 | 0.33 | 4 |
| 4-304.0 | 0.72 | 0.9 | ----- | ----- | 0.947 | 3.72 | 1.385 | 10.61 | -10.47 | 0.91 | 3 |
| 2-121.0 | 0.475 | 3.4 | ----- | ----- | 0.087 | 4.35 | ----- | ----- | ----- | ----- | 1 |
| 2-454.0 | 2.04 | 2.0 | 0.334 | 4.15 | 0.409 | 4.15 | ----- | ----- | ----- | ----- | 4 |
| 2-537.0 | 0.139 | 1.7 | ----- | ----- | ----- | ----- | ----- | ----- | ----- | ----- | 3 |
| 4-260.1 | 0.96 | 5.4 | 0.910 | 2.40 | ----- | ----- | ----- | ----- | ----- | ----- | 3 |
| 4-257.0 | 0.56 | 1.1 | ----- | ----- | 0.226 | 3.46 | 0.701 | 10.34 | -10.50 | 0.53 | 4 |
| 2-246.0 | 0.958 | 4.8 | ----- | ----- | ----- | ----- | ----- | ----- | ----- | ----- | 3 |
| 4-229.0 | 0.80 | 1.2 | 0.626 | 3.75 | 0.342 | 4.52 | 1.19 | 10.56 | -10.49 | 0.55 | 4 |
| 4-186.0 | 1.84 | 3.6 | 0.455 | 2.84 | 0.650 | 4.32 | 375.20 | 13.08 | -10.50 | 0.95 | 2 |
| 4-154.0 | 0.48 | 1.4 | ----- | ----- | ----- | ----- | 0.815 | 10.33 | -10.42 | 0.84 | 3 |
| 2-210.0 | 0.749 | 4.1 | 0.387 | 2.12 | 0.519 | 2.52 | ----- | ----- | ----- | ----- | 4 |
| 3-258.0 | 0.520 | 2.6 | ----- | ----- | 0.188 | 3.02 | ----- | ----- | ----- | ----- | 4 |
| 4-120.0 | 0.72 | 2.9 | ----- | ----- | ----- | ----- | 1.699 | 10.90 | -10.66 | 0.32 | 4 |
| 4-284.0 | 0.961 | 2.5 | 0.290 | 3.11 | ----- | ----- | ----- | ----- | ----- | ----- | 2 |
| 4-235.0 | 0.961 | 1.1 | 0.208 | 3.78 | ----- | ----- | ----- | ----- | ----- | ----- | 4 |
| 4-131.0 | 0.72 | 1.8 | ----- | ----- | 0.558 | 3.10 | 3.471 | 10.90 | -10.36 | 0.77 | 4 |
| 2-264.2 | 0.478 | 10.7 | 0.212 | 3.97 | 0.087 | 3.67 | ----- | ----- | ----- | ----- | 3 |
| 2-256.0 | 1.24 | 3.2 | 0.275 | 3.94 | 0.600 | 3.68 | ----- | ----- | ----- | ----- | 3 |
| 3-229.0 | 0.76 | 1.5 | 0.175 | 3.24 | ----- | ----- | ----- | ----- | ----- | ----- | 3 |
| 2-514.0 | 0.752 | 2.6 | 0.252 | 3.42 | 0.174 | 3.90 | ----- | ----- | ----- | ----- | 3 |
| 4-109.0 | 2.08 | 1.3 | 0.219 | 3.34 | 0.227 | 4.36 | 0.162 | 10.03 | -10.82 | 0.00 | 4 |
| 3-143.0 | 0.477 | 2.1 | 0.589 | 3.15 | 0.250 | 3.09 | ----- | ----- | ----- | ----- | 2 |
| 4-85.1 | 2.72 | 1.3 | ----- | ----- | ----- | ----- | 5.961 | 11.26 | -10.49 | 0.00 | 4 |
| 3-37.0 | 0.92 | 2.2 | 0.591 | 3.24 | ----- | ----- | ----- | ----- | ----- | ----- | 4 |
| 2-264.1 | 0.475 | 3.9 | ----- | ----- | 0.300 | 3.14 | ----- | ----- | ----- | ----- | 2 |
| 2-585.0 | 2.002 | 6.1 | ----- | ----- | ----- | ----- | ----- | ----- | ----- | ----- | 3 |
| 3-243.0 | 3.233 | 1.0 | 0.474 | 4.68 | ----- | ----- | ----- | ----- | ----- | ----- | 4 |
| 3-331.0 | 1.08 | 2.6 | 0.285 | 3.26 | ----- | ----- | ----- | ----- | ----- | ----- | 4 |
| 4-1.0 | 1.08 | 4.2 | 0.244 | 4.04 | ----- | ----- | ----- | ----- | ----- | ----- | 4 |
| 2-525.0 | 2.237 | 1.8 | 0.519 | 3.28 | 0.809 | 3.83 | ----- | ----- | ----- | ----- | 4 |
| 433.0 | 0.80 | 3.0 | 0.769 | 2.93 | 0.222 | 3.62 | 0.081 | 10.83 | -11.92 | 0.00 | 1 |
| 3-386.1 | 0.474 | 4.0 | 0.665 | 2.93 | 0.478 | 2.66 | ----- | ----- | ----- | ----- | 3 |
| 3-321.0 | 0.678 | 3.3 | ----- | ----- | ----- | ----- | ----- | ----- | ----- | ----- | 1 |
| 2-661.0 | 0.816 | 3.5 | 0.603 | 2.89 | ----- | ----- | ----- | ----- | ----- | ----- | 4 |
| 3-203.0 | 0.319 | 1.6 | 0.576 | 3.57 | ----- | ----- | ----- | ----- | ----- | ----- | 4 |
| 3-259.0 | 0.56 | 4.3 | ----- | ----- | 0.564 | 2.34 | ----- | ----- | ----- | ----- | 4 |
| 2-404.0 | 0.199 | 4.0 | 0.482 | 2.43 | 0.352 | 2.59 | ----- | ----- | ----- | ----- | 3 |
| 3-550.1 | 2.775 | 2.0 | 0.389 | 4.03 | ----- | ----- | ----- | ----- | ----- | ----- | 4 |
| 2-762.0 | 0.44 | 2.3 | ----- | ----- | 0.366 | 2.93 | ----- | ----- | ----- | ----- | 4 |
| 3-174.0 | 0.089 | 1.9 | ----- | ----- | 0.641 | 2.82 | ----- | ----- | ----- | ----- | 4 |
| 3-659.1 | 0.299 | 2.0 | 0.485 | 3.37 | 0.306 | 3.09 | ----- | ----- | ----- | ----- | 2 |
| 2-652.1 | 0.557 | 3.5 | ----- | ----- | 0.352 | 3.34 | ----- | ----- | ----- | ----- | 2 |
| 2-531.0 | 0.96 | 2.0 | 0.188 | 2.30 | 0.170 | 3.41 | ----- | ----- | ----- | ----- | 3 |
| 2-702.0 | 0.557 | 2.7 | 0.318 | 3.59 | ----- | ----- | ----- | ----- | ----- | ----- | 3 |
| 3-777.1 | 0.456 | 2.6 | ----- | ----- | ----- | ----- | ----- | ----- | ----- | ----- | 4 |
| 3-696.0 | 0.401 | 1.1 | 0.198 | 3.37 | 0.184 | 3.30 | ----- | ----- | ----- | ----- | 4 |
| 2-637.0 | 3.368 | 0.9 | 0.282 | 5.55 | ----- | ----- | ----- | ----- | ----- | ----- | 4 |
| 2-809.0 | 0.498 | 2.5 | ----- | ----- | 0.398 | 2.54 | ----- | ----- | ----- | ----- | 3 |
| 2-834.0 | 1.72 | 2.0 | ----- | ----- | ----- | ----- | ----- | ----- | ----- | ----- | 4 |
| 3-886.0 | 1.24 | 1.4 | 0.768 | 4.03 | ----- | ----- | ----- | ----- | ----- | ----- | 4 |
| 2-561.2 | 2.489 | 1.1 | ----- | ----- | ----- | ----- | ----- | ----- | ----- | ----- | 4 |
| 2-643.0 | 2.991 | 1.3 | ----- | ----- | ----- | ----- | ----- | ----- | ----- | ----- | 4 |
| 3-551.0 | 0.559 | 2.5 | 0.761 | 3.34 | 0.277 | 3.24 | ----- | ----- | ----- | ----- | 3 |
| 2-901.1 | 3.181 | 1.5 | ----- | ----- | ----- | ----- | ----- | ----- | ----- | ----- | 4 |
| 2-950.0 | 0.517 | 2.3 | ----- | ----- | ----- | ----- | ----- | ----- | ----- | ----- | 3 |
| 3-350.1 | 0.642 | 3.9 | 0.435 | 2.35 | 0.272 | 2.69 | ----- | ----- | ----- | ----- | 2 |
| 2-860.0 | 0.849 | 2.9 | 0.278 | 2.54 | 0.302 | 3.10 | ----- | ----- | ----- | ----- | 3 |
| 2-824.0 | 2.419 | 1.4 | ----- | ----- | ----- | ----- | ----- | ----- | ----- | ----- | 4 |
| 3-118.1 | 2.232 | 1.2 | 0.227 | 3.78 | ----- | ----- | ----- | ----- | ----- | ----- | 4 |
| 3-786.0 | 1.60 | 4.0 | 0.713 | 3.19 | ----- | ----- | ----- | ----- | ----- | ----- | 4 |
| 3-743.0 | 1.64 | 1.4 | 0.387 | 3.41 | ----- | ----- | ----- | ----- | ----- | ----- | 4 |
| 3-132.0 | 0.56 | 1.7 | 0.202 | 3.12 | ----- | ----- | ----- | ----- | ----- | ----- | 4 |
| 2-903.0 | 2.233 | 1.4 | 0.550 | 3.14 | 0.366 | 4.16 | ----- | ----- | ----- | ----- | 4 |
| 3-180.2 | 0.280 | 1.2 | ----- | ----- | 0.224 | 2.97 | ----- | ----- | ----- | ----- | 4 |
| 3-266.0 | 0.72 | 1.3 | ----- | ----- | 0.180 | 3.89 | ----- | ----- | ----- | ----- | 1 |
| 3-180.0 | 0.37 | 4.8 | ----- | ----- | ----- | ----- | ----- | ----- | ----- | ----- | 4 |
| 2-1023.1 | 0.564 | 3.3 | ----- | ----- | 0.655 | 3.28 | ----- | ----- | ----- | ----- | 3 |
| 2-982.0 | 1.148 | 4.5 | 0.535 | 2.48 | ----- | ----- | ----- | ----- | ----- | ----- | 4 |
| 2-1018.0 | 0.559 | 2.6 | ----- | ----- | 0.585 | 3.51 | ----- | ----- | ----- | ----- | 4 |
| 3-486.0 | 0.79 | 3.5 | 0.321 | 3.32 | 0.465 | 3.98 | ----- | ----- | ----- | ----- | 3 |
| 3-443.0 | 0.95 | 5.3 | ----- | ----- | ----- | ----- | ----- | ----- | ----- | ----- | 4 |
| 3-512.0 | 4.022 | 1.2 | 0.174 | 3.91 | ----- | ----- | ----- | ----- | ----- | ----- | 4 |
| 2-906.0 | 1.08 | 2.9 | ----- | ----- | 0.378 | 3.22 | ----- | ----- | ----- | ----- | 4 |
| 3-943.0 | 0.321 | 1.8 | ----- | ----- | 0.194 | 2.95 | ----- | ----- | ----- | ----- | 3 |
| 3-815.0 | 0.76 | 2.8 | 0.336 | 4.13 | 0.329 | 4.64 | ----- | ----- | ----- | ----- | 1 |
| 3-430.1 | 1.231 | 1.9 | ----- | ----- | 0.683 | 4.69 | ----- | ----- | ----- | ----- | 2 |
| 3-610.1 | 0.517 | 6.2 | ----- | ----- | ----- | ----- | ----- | ----- | ----- | ----- | 2 |
| 3-355.0 | 1.28 | 2.2 | ----- | ----- | 0.207 | 4.30 | ----- | ----- | ----- | ----- | 2 |
| 3-404.0 | 0.52 | 4.3 | 0.198 | 2.86 | ----- | ----- | ----- | ----- | ----- | ----- | 3 |
| 3-773.0 | 0.561 | 3.4 | ----- | ----- | 0.460 | 2.38 | ----- | ----- | ----- | ----- | 4 |
| 3-400.1 | 0.473 | 4.4 | 0.75 | 3.50 | ----- | ----- | ----- | ----- | ----- | ----- | 4 |
| 3-221.1 | 0.952 | 4.4 | ----- | ----- | 0.602 | 2.80 | ----- | ----- | ----- | ----- | 4 |
| 3-405.1 | 0.319 | 2.0 | 0.255 | 3.13 | 0.211 | 3.37 | ----- | ----- | ----- | ----- | 4 |
| 3-957.0 | 1.02 | 1.5 | 0.183 | 3.0 | 0.547 | 3.79 | ----- | ----- | ----- | ----- | 4 |
| 3-863.0 | 0.681 | 1.5 | ----- | ----- | 0.210 | 2.80 | ----- | ----- | ----- | ----- | 4 |
| 3-534.0 | 0.32 | 5.0 | ----- | ----- | 0.192 | 2.69 | ----- | ----- | ----- | ----- | 4 |
| 3-853.1 | 3.88 | 0.5 | ----- | ----- | ----- | ----- | ----- | ----- | ----- | ----- | 4 |
| 3-908.0 | 0.76 | 5.5 | ----- | ----- | ----- | ----- | ----- | ----- | ----- | ----- | 4 |
| 3-958.0 | 0.92 | 0.8 | ----- | ----- | ----- | ----- | ----- | ----- | ----- | ----- | 3 |
| 3-875.0 | 2.04 | 2.8 | 0.643 | 3.12 | ----- | ----- | ----- | ----- | ----- | ----- | 1 |
| 3-790.1 | 0.562 | 2.6 | 0.226 | 3.31 | 0.124 | 3.82 | ----- | ----- | ----- | ----- | 1 |

[1] ID is from Williams et al. (1996) catalog, where numbers have been truncated to one decimal place after the primary designation. $r_1$ denotes the first-moment galaxy radius from Williams et al. (1996), converted to kiloparsecs. Star formation rate (SFR) is in units of solar masses per year. Bolometric luminosity ($L_{bol}$) is in units of solar bolometric luminosity. The parameter "frac" is the fraction of the bolometric luminosity radiated in the mid- and far-infrared (see TWSL). Spectral type is taken from Fernández-Soto et al. (1999) and is coded as follows: 1 = Elliptical galaxy, 2 = Sbc galaxy, 3 = Scd galaxy, 4 = Irregular galaxy.

FIGURE CAPTIONS

FIGURE 1. – Sample galaxies from the HDFN illustrating the range in morphological asymmetries. Figure 1*a* shows the images covering the rest-frame visible emission of the galaxies 4-916.0 ($z = 0.16$), 4-558.0 ($z = 0.48$), 3-430.1 ($z = 1.231$) and 4-660.0 ($z = 2.32$). The images are 3" square. The first two galaxies are shown in the WFPC2 (F606W + F814W) image, while the last two galaxies are shown in the NICMOS (F110W + F160W) image. Figure 1*b* shows the brightness profiles of the galaxies in these images, averaged over 2 - 5 pixels along their major axes, along with their measured asymmetry and concentration parameter values. Note the correspondence of the values of the asymmetry parameter to the appearance of the brightness profiles.

FIGURE 2. -- Comparison of the rest-frame ultraviolet and rest-frame visible asymmetry and concentration parameters for the sample galaxies, for each object in which both could be measured. Correlation coefficients for these comparisons are listed in Table 2. Object symbols are based on the classification of their spectral energy distributions by Fernández-Soto et al. (1999), with triangles denoting objects best fitted by an Elliptical galaxy spectrum, squares denoting objects best fitted by an Sbc galaxy spectrum, closed circles denoting objects best fitted by an Scd galaxy spectrum, and open circles denoting objects best fitted by an Irregular galaxy spectrum.

FIGURE 3. – Comparison of galaxy bolometric luminosities and star formation rates. Symbol sizes are based on the set of spectral templates used by TWSL to fit the object spectral energy distributions. Larger circles represent galaxies best fitted by "colder" templates having lower rest-frame ultraviolet flux and older stellar populations.

FIGURE 4. – (*a*) Comparison of galaxy rest-frame visible asymmetry values and redshifts. Symbols are the same as in Figure 2, but with crosses representing the measurements of C2000 for the nearby galaxy sample of Frei et al. (1996). (*b*) Mean values of data in (*a*), binned for five sub-samples of approximately equal size. Error bars represent $1\sigma$ values of each quantity within the sub-samples.

FIGURE 5. – Same as Figure 4*b*, but after converting redshift to age under the adopted cosmology.

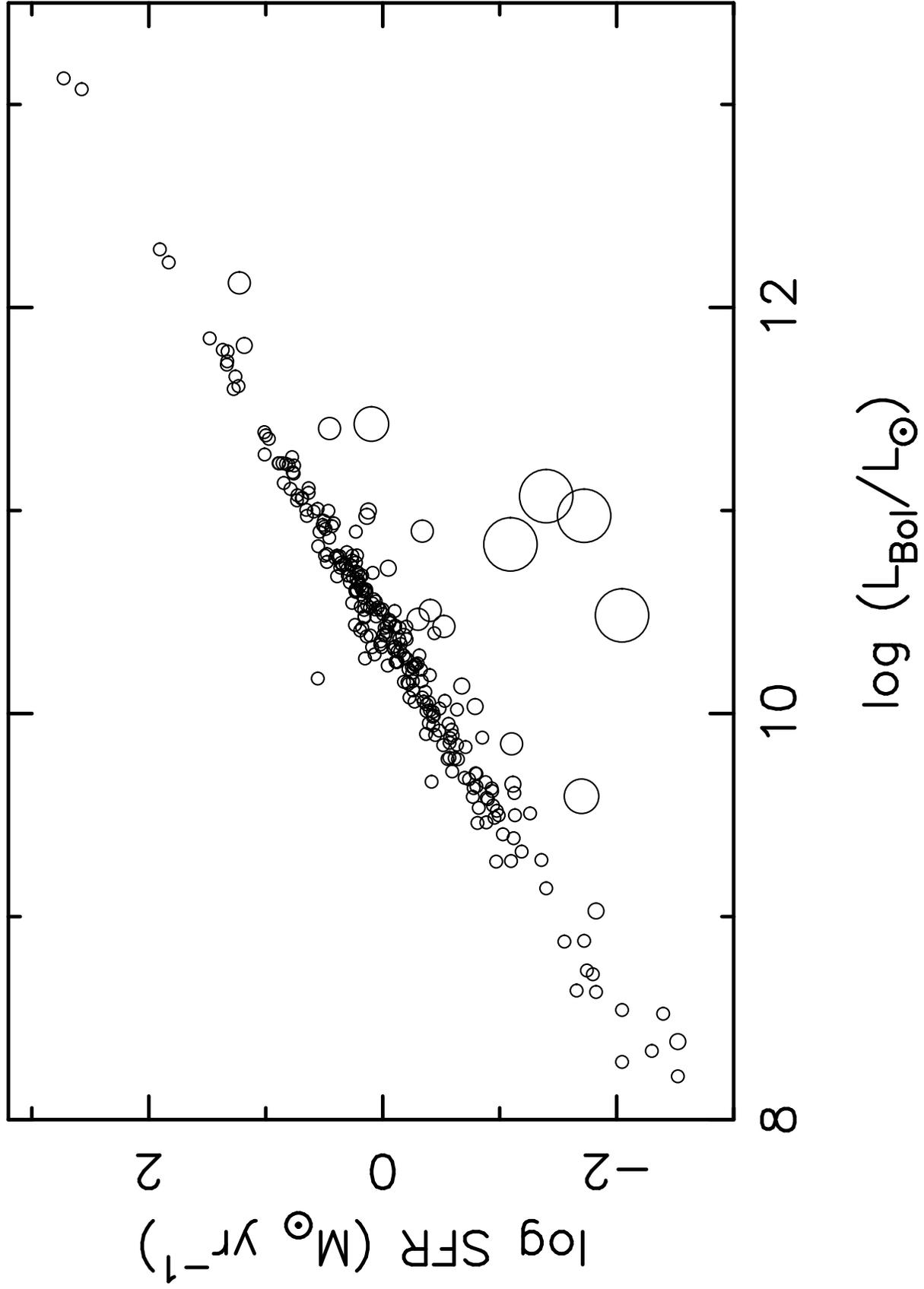

TABLE 2

CORRELATION MATRIX[1]

| Parameter | z | $r_1$ | A(UV) | C(UV) | A(Vis) | C(Vis) | SFR | log $L_{bol}$ | log(SFR/L) | frac |
|---|---|---|---|---|---|---|---|---|---|---|
| z | 1 | -0.044 | -0.010 | 0.351 | 0.557 | 0.211 | 0.520 | 0.500 | 0.155 | -0.528 |
| $r_1$ | -0.044 | 1 | 0.088 | -0.467 | -0.122 | -0.323 | 0.447 | 0.540 | 0.133 | 0.120 |
| A(UV) | -0.010 | 0.088 | 1 | -0.130 | 0.436 | -0.216 | 0.013 | 0.081 | 0.253 | 0.121 |
| C(UV) | 0.351 | -0.467 | -0.130 | 1 | 0.029 | 0.630 | 0.017 | -0.065 | -0.061 | -0.410 |
| A(Vis) | 0.557 | -0.122 | 0.436 | 0.029 | 1 | -0.216 | -0.002 | -0.221 | 0.135 | -0.024 |
| C(Vis) | 0.211 | -0.323 | -0.216 | 0.630 | -0.216 | 1 | 0.089 | 0.118 | -0.165 | -0.136 |
| SFR | 0.520 | 0.447 | 0.013 | 0.017 | -0.002 | 0.089 | 1 | 0.923 | 0.471 | 0.134 |
| log $L_{bol}$ | 0.500 | 0.540 | 0.081 | -0.065 | -0.221 | 0.118 | 0.923 | 1 | 0.238 | 0.041 |
| log(SFR/L) | 0.155 | 0.133 | 0.253 | -0.061 | 0.135 | -0.165 | 0.471 | 0.238 | 1 | 0.329 |
| frac | -0.528 | 0.120 | 0.121 | -0.410 | -0.024 | -0.136 | 0.134 | 0.041 | 0.329 | 1 |

[1] Values are the Spearman rank test correlation coeffiecient

TABLE 3

PRINCIPAL COMPONENTS ANALYSIS OF CORRELATION MATRIX[1]

| eigenvector | z | $r_1$ | A(UV) | C(UV) | A(Vis) | C(Vis) | SFR | log $L_{bol}$ | log(SFR/L) | frac |
|---|---|---|---|---|---|---|---|---|---|---|
| 1. | -0.447 | 0.827 | 0.368 | -0.945 | -0.164 | -0.819 | 0.460 | 0.431 | 0.660 | 0.727 |
| 2. | 0.482 | 0.373 | -0.665 | 0.139 | -0.570 | 0.360 | 0.836 | 0.866 | -0.032 | -0.348 |
| 3. | -0.740 | -0.045 | -0.415 | 0.014 | -0.763 | 0.378 | -0.224 | -0.145 | -0.199 | 0.485 |
| 4. | -0.058 | 0.374 | 0.118 | -0.184 | 0.006 | -0.103 | -0.143 | 0.074 | -0.702 | -0.190 |

[1] Values are coefficients of the linear addition of parameter values. Eigenvalues of the individual eigenvectors are 3.97 (eigenvector 1), 2.86 (eigenvector 2), 1.79 (eigenvector 3), 0.76 (eigenvector 4).

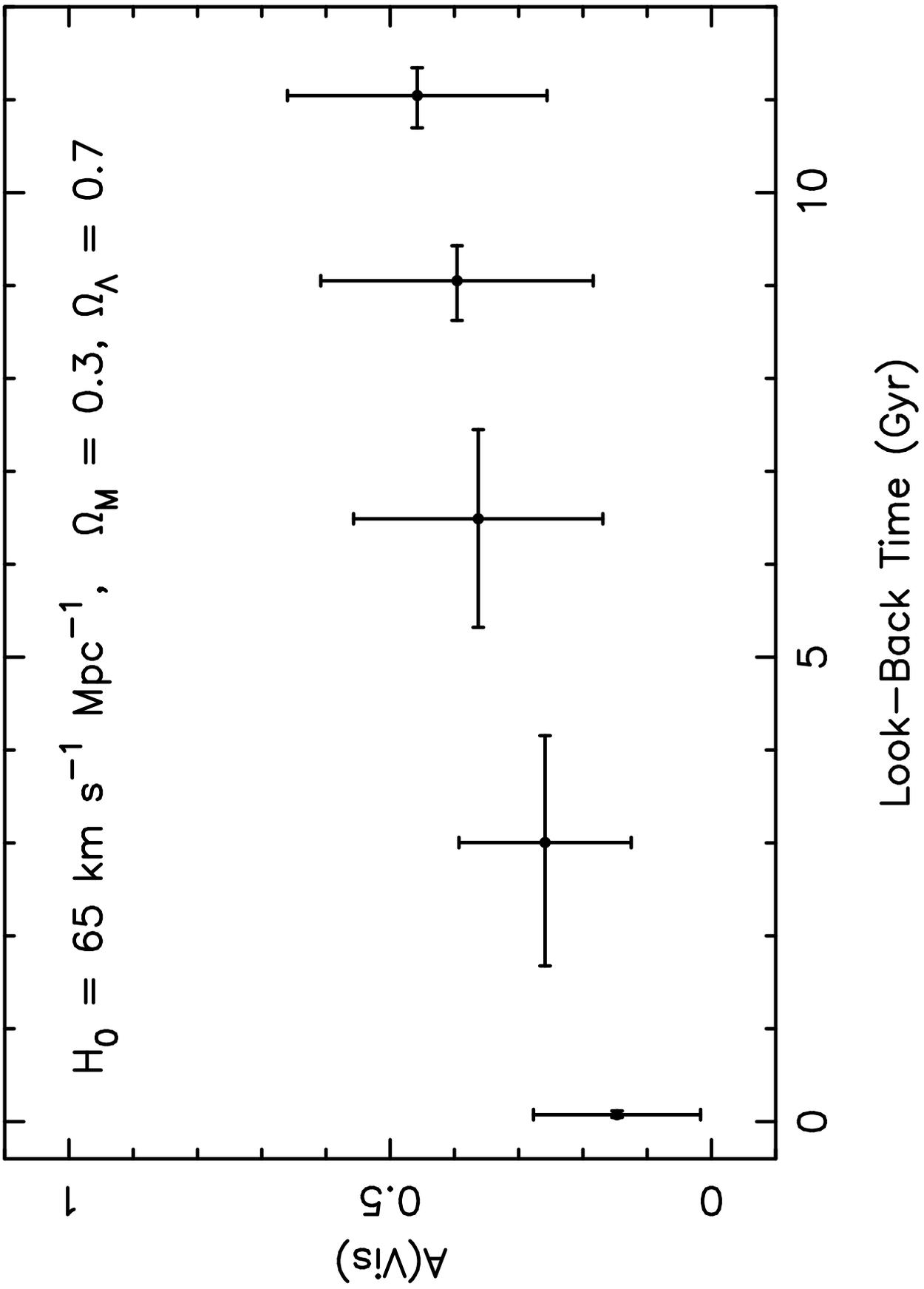